\documentclass[]{spie}  %>>> use for US letter paper
\usepackage{amsfonts}
\usepackage{amsmath}

\usepackage{multirow} 
\usepackage{amsmath,amsfonts,amssymb}
\usepackage{graphicx}
\usepackage[colorlinks=true, allcolors=blue]{hyperref}
\usepackage{comment}

\title{Disentangled Multimodal Brain MR Image Translation \\via Transformer-based Modality Infuser}

\author[a,b]{Jihoon Cho}
\author[a]{Xiaofeng Liu}
\author[a]{Fangxu Xing}
\author[a]{Jinsong Ouyang}
\author[a]{Georges El Fakhri}
\author[b]{Jinah Park}
\author[a]{Jonghye Woo}

\affil[a]{{Gordon Center for Medical Imaging, Massachusetts
General Hospital and Harvard Medical School, Boston, MA 02114 USA}}
\affil[b]{Korea Advanced Institute of Science and Technology, 291 Daehak-ro, Yuseong-gu, Daejeon 34141, Republic of Korea}

\begin{document} 
\maketitle

\vspace{+5pt}
\begin{abstract}
Multimodal Magnetic Resonance (MR) Imaging plays a crucial role in disease diagnosis due to its ability to provide complementary information by analyzing a relationship between multimodal images on the same subject. Acquiring all MR modalities, however, can be expensive, and, during a scanning session, certain MR images may be missed depending on the study protocol. The typical solution would be to synthesize the missing modalities from the acquired images such as using generative adversarial networks (GANs). Yet, GANs constructed with convolutional neural networks (CNNs) are likely to suffer from a lack of global relationships and mechanisms to condition the desired modality. To address this, in this work, we propose a transformer-based modality infuser designed to synthesize multimodal brain MR images. In our method, we extract modality-agnostic features from the encoder and then transform them into modality-specific features using the modality infuser. Furthermore, the modality infuser captures long-range relationships among all brain structures, leading to the generation of more realistic images. We carried out experiments on the BraTS 2018 dataset, translating between four MR modalities, and our experimental results demonstrate the superiority of our proposed method in terms of synthesis quality. In addition, we conducted experiments on a brain tumor segmentation task and different conditioning methods.

\vspace{+5pt}
\end{abstract}


\vspace{+5pt}
\section{Introduction}\vspace{+5pt}

Magnetic Resonance Imaging (MRI) is a widely used imaging modality for clinical diagnosis and treatment, offering efficient extraction of abundant information from soft tissue regions. Multimodal MR images provide complementary information on pathology or tissue morphology, leading to more precise and reliable results~\cite{cho2023mmTrans, cho2023HFTrans}. However, obtaining a sufficiently abundant set of multimodal MRI scans can be challenging, due to time constraints, limited imaging equipment availability, or poorly planned study protocols. To address this issue, in recent years, various image synthesis methods using generative adversarial networks (GANs) have been proposed to synthesize missing MR images. For example, CycleGAN~\cite{zhu2017unpaired} introduced cycle consistency loss to synthesize cross-modal images from unpaired data. Pix2Pix~\cite{isola2017image}, on the other hand, proposed an image translation network for pairwise data. They showed excellent performance in synthesizing multi-contrast MRI~\cite{dar2019image}, but due to the limitations of its one-way translation approach, it is not well-suited for synthesizing multiple modalities. In the case of TC-MGAN~\cite{xin2020multi}, its 1-to-3 network makes use of a pre-trained tumor segmentation network to generate T1-weighted (T1) MRI, post-contrast T1-weighted (T1ce) MRI, and T2-FLAIR (FLAIR) MRI from T2-weighted (T2) MRI. Recently, UCD-GAN~\cite{liu2021unified} proposed a disentanglement framework, capable of translating all pairs of modality using a single network.

While the aforementioned approaches have demonstrated promising results in image synthesis, they still suffer from fundamental limitations associated with CNN models. These limitations include the lack of ability to capture long-range dependencies essential for understanding the entire structure and inefficient conditioning mechanisms for synthesizing specific modalities. To overcome these limitations, in this work, we propose a disentangled multimodal MR image translation framework using a transformer-based modality infuser. Inspired by the approach in Liu et al~\cite{liu2021unified}, our framework disentangles modality-invariant features and modality-specific features. The modality infuser takes the modality-invariant features from the CNN encoder as input and converts them into modality-specific features. Specifically, we construct modality infusers using transformer layers, incorporating modality encoding—similar to positional encoding~\cite{vaswani2017attention}—in each transformer layer. This approach offers several advantages, including the ability to capture the entire brain structure through self-attention and effectively disentangle features by clearly defining distinct roles within the network.

\begin{comment}
\begin{figure}[t]
\begin{center}
\includegraphics[width=0.8\linewidth]{fig_transformer.png}
\end{center} 
\caption{Structure of the modality infuser with consecutive modality encoding} 
\label{fig:Transforemr}
\end{figure} 
 
\vspace{+5pt}
\section{Methods}\vspace{+5pt}
\end{comment}

\begin{figure}[t]
\begin{center}
\includegraphics[width=1.0\linewidth]{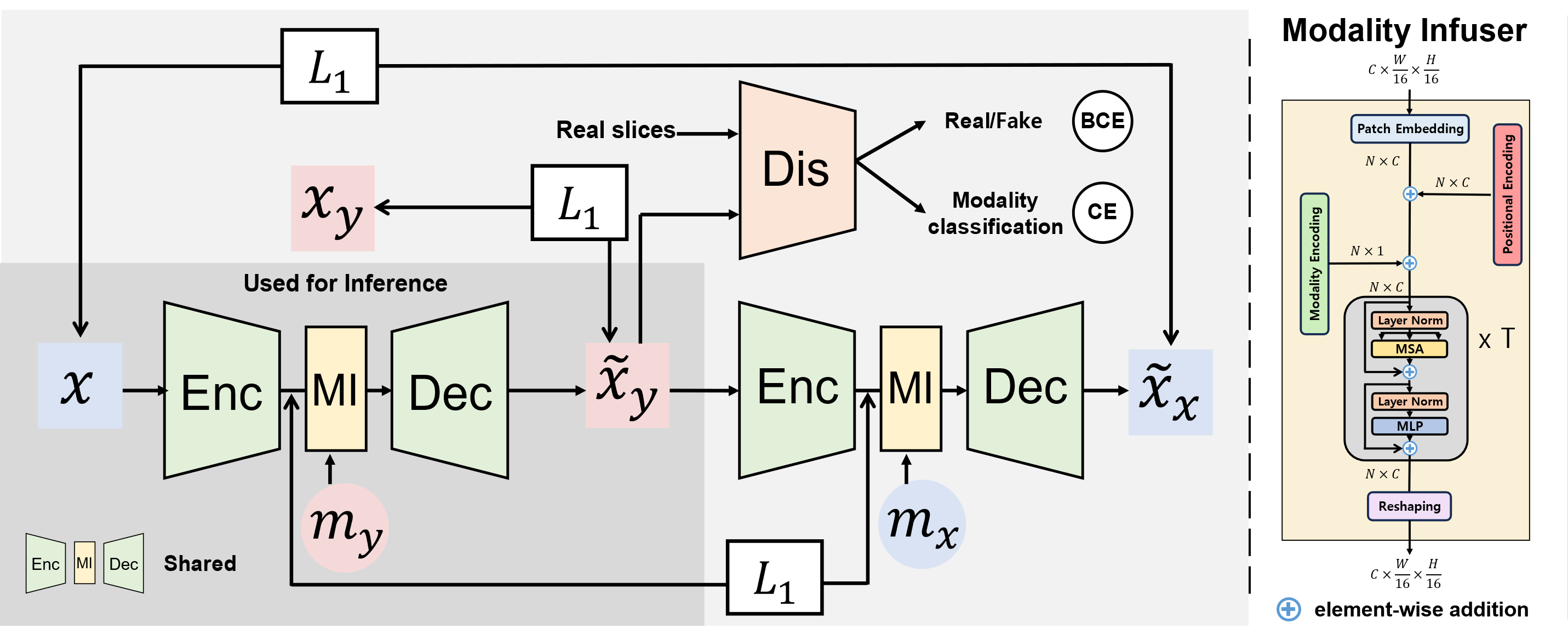}
\end{center} 
\caption{Illustration of our framework for translating MR images (left) and structure of the modality infuser (right). Our framework consists of CNN Encoder (Enc), transformer-based modality infuser (MI), CNN Decoder (Dec), and CNN discriminator (Dis).} 
\label{fig:overall}
\end{figure} 
 
\vspace{+5pt}
\section{Method}%\vspace{+5pt}

When a set of $M$ MR images is provided for a co-registered brain area, a sample $x \in \mathbb{R}^{1\times{W}\times{H}}$ with modality $m_x$ is accompanied by $M-1$ paired samples correspoding to the other modalities. The goal of this work is to synthesize $M-1$ MR sequences $\tilde{x}_y$ with modality $m_y \in [m_1, m_2, ..., m_M] - m_x$ from a given image $x$, irrespective of the specific modality $m_x$.
To achieve the synthesis of all modality pairs using a single network, we adopt the training concept of modality disentanglement, as proposed in~\cite{liu2021unified}. First, we extract the features $f_x \in \mathbb{R}^{C\times{W\over 16}\times{H\over 16}}$ from $x$ through the CNN encoder and then force the features to be modality-agnostic by applying the loss of modality disentanglement $L_{disen} = |f_x - f_{\tilde{x}_y}|$~\cite{liu2021unified,liu2019feature,mathieu2016disentangling}. 
Next, we convert the features $f_x$ to become modality-informative for the generation of target modality images at the CNN decoder. The modality infuser takes modality-invariant features and embeds them into 1D sequences $z_0 \in \mathbb{R}^{N(={W\over 16}\times{H\over 16})\times C}$ by incorporating positional encoding. The addition of positional encoding assigns unique positional representations, aiding in understanding global structure~\cite{dosovitskiy2020image}. Then sequences $z_0$ are processed $T$ times consecutively by the self-attention layers with the addition of the modality encoding ($ME \in \mathbb{R}^{1 \times C} $), i.e., $ME(m_y,2i) = sin( {{m_y}\over {10000^{{2i}\over {C}}}})$ and $ME(m_y,2i+1) = cos( {{m_y}\over {10000^{{2i+1}\over C}}}),$ where $i$ denotes channel dimension. At the end of the modality infuser, the sequences $z_L\in \mathbb{R}^{N\times C}$ reshape to the original shape of the input features.

Finally, modality-conditioned features are used to generate the target modality through the CNN decoder, and the resulting generated image $\tilde{x}_y$ is compared with the ground truth $x_y$ using the L1 loss, $L_{rec}$. In addition, we apply the cycle-consistency loss $L_{cyc}$ to better maintain the shape structures~\cite{zhu2017unpaired}. However, the same network is recalled twice to generate the input image $\tilde{x}_x$ as the final output, because the proposed network is able to effectively translate all of the modality pairs~\cite{liu2021unified}. 
For adversarial training, we introduce a discriminator $D$ responsible for discerning real inputs and determining their respective modalities. The inclusion of an auxiliary modality classification task is beneficial in achieving two key objectives: enabling training with a single discriminator and facilitating the generation of probability distributions over the modalities~\cite{odena2017conditional}. Therefore, the discriminator is optimized by two kinds of losses---the binary cross-entropy loss for identifying real slices $L_{adv}=\mathbb{E}_{x}[logD(x)] + \mathbb{E}_{\tilde{x}_y}[1-logD(\tilde{x}_y)]$ and the auxiliary cross-entropy loss for modality classification $L_{aux}=\mathbb{E}_{x,m_x}[-logD_{aux}(x,m_x)] + \mathbb{E}_{\tilde{x}_y,m_y}[-logD_{aux}(\tilde{x}_y,m_y)]$. For the image translator, it is trained by optimizing $L_G=\alpha L_{rec}+\beta L_{disen}+\gamma L_{cyc}+\lambda_1 L_{adv}+\lambda_2 L_{aux}$, where $\alpha$, $\beta$, $\gamma$, $\lambda_1$, $\lambda_2$ are weightings for each loss item. An overview of the proposed training framework is shown in Fig.~\ref{fig:overall}.

\begin{figure}[t]
\begin{center}
\includegraphics[width=1.0\linewidth]{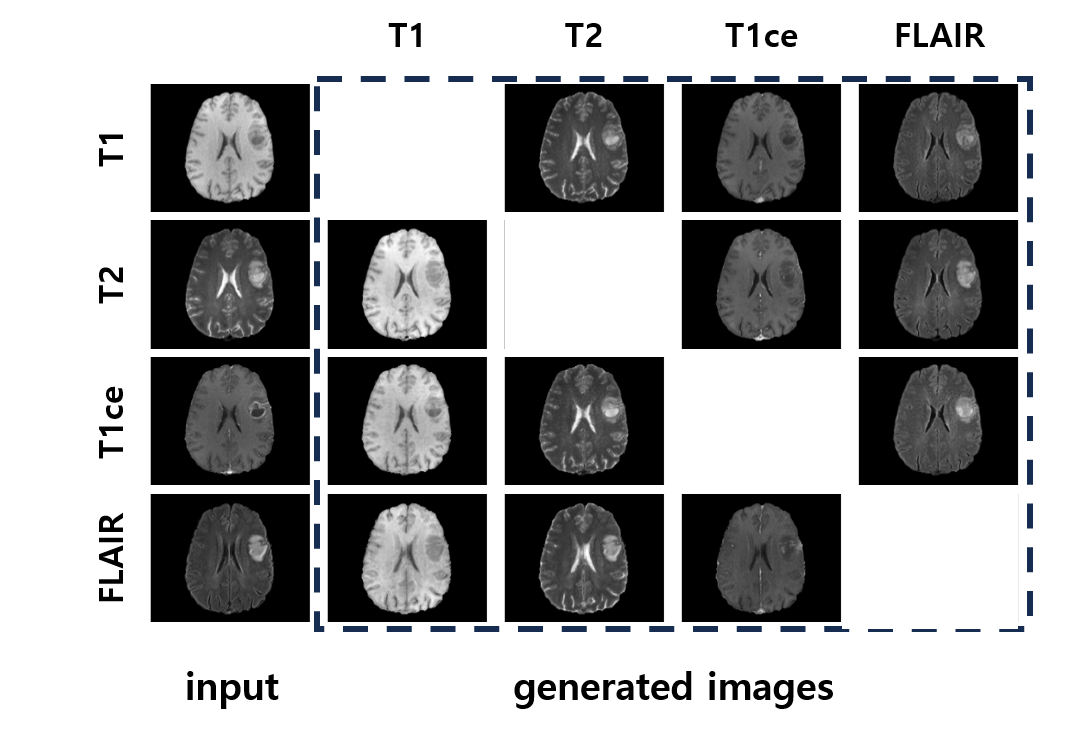}
\end{center} 
\caption{Multimodal MR image synthesis results using our framework. The MR images of the first column are translated into the other modalities.}% all modalities including self-translation ($m_x = m_y$).} 
\label{fig:results}
\end{figure}

\vspace{+5pt}
\section{Results and Discussion}\vspace{+5pt}

%\subsection{Experiments Settings}
We experimented with the BraTS 2018 dataset to validate the effectiveness of our method. BraTS 2018 consists of a total of 285 subjects, each with four MR modalities: T1, T2, T1ce, and FLAIR MRI. To evaluate our method with the previous method by~\cite{liu2021unified}, we follow the identical configuration used in that work~\cite{liu2021unified}. Specifically, we partitioned the dataset into four subsets: 80 subjects for training, 20 subjects for validation, 100 subjects for testing, and additional 100 subjects for tumor segmentation experiments. The images have a size of $240 \times 240 \times 155$, and, to standardize the intensity range, we linearly scaled the original intensities to $[-1, 1]$. For training, we used 2D axial slices, discarding those with fewer than 2000 brain pixels. 

Our framework was implemented using PyTorch, and the networks were trained for 100 epochs on an NVIDIA RTX 3090 GPU with a batch size of 24. We used Adam optimizer with learning rates of 0.0001 and 0.00001 for the generator and the discriminator, respectively. We set the hyper-parameters as $\alpha = 10,\, \beta = 1,\, \gamma = 1,\, \lambda_1 = 0.25,\, \lambda_2 = 0.25,\,$ and the number of transformer layers $T=4$.

\subsection{Quantitative Evaluations}
We evaluated the 12 synthesized results for each subject, considering all kinds of modality mapping, using four metrics including mean absolute error (L1), peak signal-to-noise ratio (PSNR), structural similarity measure (SSIM), and multi-scale SSIM (MS-SSIM). Evaluation results are shown in Table ~\ref{table:results}. The standard deviation was determined by three separate random experiments. The proposed method outperforms the previous method in all metrics. It has demonstrated the ability to overcome the fundamental limitation of CNN models, by infusing modality information through the transformer layers. In particular, the single-mode encoding applied in the only first transformer layer showed improved synthesis quality compared with consecutive modality encoding, which guarantees that the simple structure of ME can infuse sufficient modality information. The synthesized results from the proposed framework are shown in Fig.~\ref{fig:results}.

\begin{table}[t]
\centering 
\caption{Numerical comparisons with the previous method. The best results are \textbf{bold}.} \vspace{+8pt}
\resizebox{1\linewidth}{!}{
\begin{tabular}{l|c|c|c|c}
\hline
Methods& ~~L1 (x1000) $\downarrow$~~ & ~~~PSNR $\uparrow$~~~ & ~~~SSIM $\uparrow$~~~ & ~~~MS-SSIM $\uparrow$~~~\\\hline\hline
UCD-GAN~\cite{liu2021unified}  &  {16.42}$\pm$0.082 & {28.10}$\pm$0.008  &  {0.9013}$\pm$0.0005  & {0.9166}$\pm$0.0005  \\ \hline
Ours w/ consecutive ME   & {14.70}$\pm$0.038 & {29.12}$\pm$0.015  &  {0.9281}$\pm$0.0002  & {0.9408}$\pm$0.0002  \\
Ours w/ single ME  & \textbf{14.62}$\pm$0.043 & \textbf{29.15}$\pm$0.029  &  \textbf{0.9287}$\pm$0.0004  & \textbf{0.9414}$\pm$0.0004   \\ \hline
\end{tabular}} 
\label{table:results}  
\end{table}

\subsection{Tumor Segmentation Results}
Synthesizing missing modalities can be useful for analyzing disease diagnosis, such as brain tumor segmentation, which requires all modalities for accurate prediction. In addition, the synthesized images can be used as training data to improve the accuracy of segmentation. To demonstrate the capability, we assumed that only FLAIR MRI is provided, and then trained the U-Net~\cite{unet} to predict the brain tumor region with the synthesized MR images. Table~\ref{table:results_seg} shows the segmentation results. Training the segmentation network with synthesized T1, T2, and T1ce MR images by our method showed better results than training with real FLAIR images alone, while the previous synthesis method showed similar segmentation results. Our proposed method demonstrated the ability to capture the unique characteristics of each modality using a transformer-based modality infuser and showed the possible improvement in prediction performance when the training data have missing modalities by synthesizing the missing modalities instead of discarding the data.

\begin{table}[b]
\centering 
\caption{Brain tumor segmentation results trained with synthesized MR images. The standard deviation was determined by three separate random experiments. The best results are \textbf{bold}.} \vspace{+8pt}
\resizebox{0.6\linewidth}{!}{
\begin{tabular}{l|c|c|c}
\hline
\multirow{2}{*}{Method} & \multicolumn{2}{c|}{Modalities}  & ~~~\multirow{2}{*}{Dice Score $\uparrow$} \\ \cline{2-3} & Real & Synthesized  \\ \hline \hline
Real  &  FLAIR & - &  0.7478$\pm$0.0005  \\ 
UCD-GAN~\cite{liu2021unified}  &  FLAIR & T1, T2, T1ce  &  0.7420$\pm$0.0075    \\
Ours   &  FLAIR & T1, T2, T1ce  &  \textbf{0.7579}$\pm$0.0002    \\ \hline
% Real - All  &  T1, T2, T1ce, FLAIR & -  &  0.7976$\pm$0.0005    \\ \hline
\end{tabular}} 
\label{table:results_seg}  
\end{table}

\subsection{Modality conditioning method}
We proposed the novel conditioning method using the fixed value of each modality as the modality encoding within the modality infuser constructed with the transformer. We conducted an ablation study of the modality encoding methods that include the learnable parameters rather than fixed values. As shown in Table~\ref{table:pe}, the learnable parameter demonstrated difficulty in training the modality conditioning ability, even when the high reconstruction loss $L_{rec}$ is applied. This demonstrates the importance of providing distinctly different values for modality encoding, which can be achieved by the proposed method.

\begin{figure}[h]
\begin{center}
\includegraphics[width=0.6\linewidth]{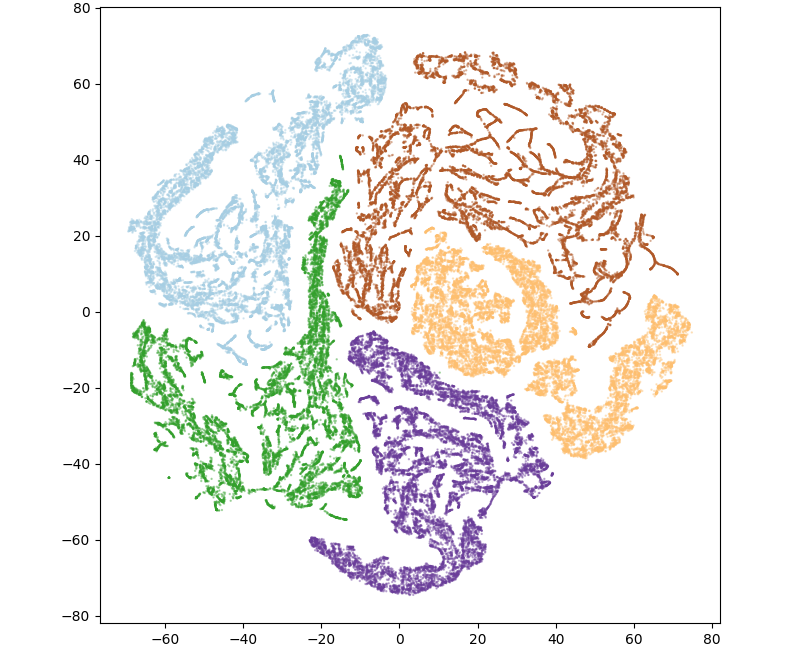}
\end{center} 
\caption{Feature visualization results: Brown color depicts modality-agnostic features extracted from the CNN encoder, while other colors represent conditioned features through the modality infuser (T1: blue, T2: green, T1ce: yellow, and FLAIR: purple.)}
\label{fig:feature}
\end{figure}

In addition, we visualized the modality-agnostic features extracted from the CNN encoder and the modality-specific features transformed by the modality infuser as shown in Fig.~\ref{fig:feature}. We reduced the feature dimension to 50 using principal component analysis (PCA) and applied the t-distributed stochastic neighbor embedding (t-SNE) to visualize the features in 2D space. We observed that the conditioned features of each modality and unconditioned features are placed in each cluster; therefore, the conditioning mechanism of our proposed method has shown the capability of converting to the specific feature space of the modality.

\begin{table}[t]
\centering 
\caption{Ablation study of modality encoding method. Learnable parameter* is trained with higher reconstruction loss $L_{rec}$ with the weight of $\alpha = 50$. The best results are \textbf{bold}.} \vspace{+8pt}
\resizebox{1\linewidth}{!}{
\begin{tabular}{l|c|c|c|c}
\hline
Modality Encoding& ~~L1 (x1000) $\downarrow$~~ & ~~~PSNR $\uparrow$~~~ & ~~~SSIM $\uparrow$~~~ & ~~~MS-SSIM $\uparrow$~~~\\\hline\hline
Learnable parameter   & 23.48 & 25.01 & 0.8675 & 0.8244  \\
Learnable parameter*   & 20.95 & 26.05 & 0.8813 & 0.8547  \\
Fixed parameter  & \textbf{14.57} & \textbf{29.20}  &  \textbf{0.9292}  & \textbf{0.9421}   \\ \hline
\end{tabular}} 
\label{table:pe}  
\end{table}

\vspace{+5pt}
\section{Conclusion}\vspace{+5pt}
In this work, we proposed a synthesis framework for translating MR images between different modalities. In particular, we developed a modality infuser that allows global relationships to be obtained through a transformer and converts modality-specific information through modality encoding.
Our experimental results demonstrated that our framework was able to translate MR images accurately using a single network, outperforming the previous CNN models. In addition, the necessity of the proposed conditioning method has been quantitatively and qualitatively confirmed.

\vspace{+5pt}
\acknowledgments % equivalent to \section*{ACKNOWLEDGMENTS}            
This work was supported by Institute for Information \& communications Technology Promotion(IITP) grant funded by the Korea government(MSIT) (No.00223446, Development of object-oriented synthetic data generation and evaluation methods) and the Technology Innovation Program (20011875, Development of AI-Based Diagnostic Technology for Medical Imaging Devices) funded by the Ministry of Trade, Industry \& Energy (MOTIE, Korea).

% References
\bibliography{main} % bibliography data in report.bib
\bibliographystyle{spiebib} % makes bibtex use spiebib.bst

\end{document}